\documentstyle[prl,aps,psfig]{revtex}

\draft

\begin{document}
\twocolumn[\hsize\textwidth\columnwidth\hsize\csname
@twocolumnfalse\endcsname

\draft

\title{
\hbox to\hsize{\large Submitted to Phys.~Lett.~B \hfil E-Print
astro-ph/9610221}
\vskip1.55cm
Cosmological Neutrino Signatures for Grand Unification
Scale Physics}

\author{G\"unter Sigl, Sangjin Lee, and David N. Schramm}
\address{Department of Astronomy \& Astrophysics,
Enrico Fermi Institute, The University of Chicago, Chicago, IL~~60637-1433\\
NASA/Fermilab Astrophysics Center,
Fermi National Accelerator Laboratory, Batavia, IL~~60510-0500}

\author{Paolo Coppi}
\address{Department of Astronomy,
Yale University, New Haven, CT 06520-8101}

\maketitle

\begin{abstract}
Physics beyond the standard model might imply the cosmological
production of particles with grand unification scale
energies. Nucleons and $\gamma$-rays from such processes are
candidates for the cosmic rays observed beyond $100\,$EeV ($10^{20}\,$eV).
Using a new particle propagation code, we calculate the neutrino
fluxes predicted by such scenarios if consistency with the
observed cosmic ray flux and the universal $\gamma$-ray
background at $1-10\,$GeV is required. Flux levels
detectable by proposed km$^3$ scale neutrino
observatories are allowed by these constraints. Bounds on
or detection of a neutrino flux above $\sim1\,$EeV would allow neutrino
astronomy to probe grand unification scale physics.
\end{abstract}

\vskip2pc]

\narrowtext

The recent detection of ultra-high energy cosmic rays (UHE CRs)
with energies above $100\,$EeV~\cite{fe1,ag1}
has prompted an intensive discussion in the literature on
the nature and origin of these particles~\cite{SSB,HVSV,ES}. At
these energies, nucleons and heavy nuclei are
subject to  photopion production on the cosmic microwave
background (CMB) (the Greisen-Zatsepin-Kuzmin
effect~\cite{GZK}), and to photodisintegration via the giant
dipole resonance~\cite{Puget}, respectively, which limits their
range to less than about $30\,$Mpc. Together with the fact that
protons above $100\,$EeV are expected to be deflected by only a
few degrees in the galactic and extragalactic magnetic
fields~\cite{SSB}, this puts severe constraints on models
explaining these events by acceleration of charged particles in
active galactic nuclei (AGN) or other known astronomical objects.

In a more speculative class of so-called ``top-down'' (TD) models,
predominantly $\gamma$-rays, electrons (positrons), and
neutrinos are produced directly at UHEs by the
cascades initiated by the decay of a supermassive elementary ``X''
particle associated with some grand unified theory (GUT), rather
than being accelerated. The X
particles are usually thought to be released from
topological defects such as cosmic strings, domain walls and
magnetic monopoles left over from GUT phase transitions.
Models of this type
predict spectra which are considerably harder and extend much
further beyond $100\,$EeV than shock acceleration spectra and
therefore appear as a natural alternative to explain the highest
energy events observed~\cite{BHS}. The predicted flux level is very model
dependent~\cite{GK}, but, in at least some of the scenarios
considered~\cite{BS}, is consistent with the observed UHE CR
fluxes.

Several signatures which might distinguish top-down scenarios
from more conventional ``bottom-up'' acceleration models have
been suggested in the literature: above 
$\simeq100$ EeV, domination of the UHE CR flux
by a hard $\gamma$-ray component  \cite{ABS}, which 
potentially produces a ``gap''~\cite{SLSB} in the UHE CR spectrum,
and at lower energies $\simeq10\,$ EeV, a $\gamma$-ray to total cosmic ray
flux ratio on the order of $10\%$~\cite{SLC}. Another potentially
important signature of TD models is their prediction of a large 
cosmological neutrino flux~\cite{BHS} which above
$100\,$EeV can dominate by far 
contributions expected from other sources such as AGN~\cite{AGN}. 
TD scenarios also predict a substantial $\gamma$-ray flux at 
lower (GeV) energies that is constrained by
the observed background levels ~\cite{Chi,SJSB,SLC,PJS}. 
In this Letter, we use a new propagation code ~\cite{Lee} to 
establish a relation between the UHE neutrino flux predictions
within TD scenarios and the constraints arising from 
the corresponding $\gamma$-ray flux predictions around $1\,$GeV
and above $\sim10.$EeV. For
km$^3$ scale neutrino observatories, prototypes of which
are AMANDA, DUMAND, Baikal, and NESTOR~\cite{detector}, we
estimate the UHE neutrino event rates predicted by a TD
scenario which explains UHE CRs and 
give the maximum event rates consistent with the $\gamma$-ray flux
constraints. We also estimate rates
for deeply penetrating showers for
fluorescence detectors such as the High Resolution Fly's
Eye (HiRes)~\cite{hires} and the Telescope Array~\cite{tel}, and for
horizontal neutrino events that might be observed in the
proposed Pierre Auger project~\cite{Cronin}.

The X particles released from topological defects via
physical processes such as collapse or annihilation could have GUT
scale masses $m_X$ up to $\sim 10^{16}\,$GeV. We assume that the X particles 
quickly decay into leptons and quarks of comparable energy.
We take the primary lepton produced in
a decay to be an electron with energy $m_X/2$. This lepton was
not included in prior calculations but can
significantly contribute to the
$\gamma$-ray flux predicted at low energies.
The quarks produced in the decays interact strongly and 
fragment into jets of hadrons typically containing  $10^4-10^5$
mesons and baryons.
Given the X particle production rate, $dn_X/dt$, the effective
production spectrum for particle species $a$ ($a=\gamma,N,e^\pm,\nu$) 
via this hadronic channel can be
written as $\Phi_a(E,t)=(dn_X/dt)(2/m_X)(dN_a/dx)$,
where $x \equiv 2E/m_X$, and $dN_a/dx$ is the relevant effective
fragmentation function. 
For the total hadronic fragmentation function $dN_h/dx$ we use
solutions of the QCD evolution equations in modified leading
logarithmic approximation which provide good fits to accelerator
data at LEP energies~\cite{detal}.
We assume that about 3\% of the total hadronic content consists of
nucleons and the rest is produced as pions and distributed equally among
the three charge states.
The effective production spectra of $\gamma$-rays, electrons, and
neutrinos are then determined from the pion decay spectra.
The X particle production rate is assumed to be
spatially uniform and in the matter-dominated era can be
parametrized as~\cite{BHS} $dn_X/dt\propto t^{-4+p}$,
where $p$ depends on the specific defect scenario.
A network of ordinary cosmic strings~\cite{BR} and
annihilation of magnetic monopole-antimonopole pairs~\cite{BS}
are represented by $p=1$, whereas $p=2$ corresponds to a constant
comoving production rate. 
As there is considerable uncertainty
in the TD model physics, we will not explicitly consider any specific
TD scenario. Rather, we parametrize a TD model 
by the typical X particle mass, $m_X,$
the X particle production rate at zero redshift,
$(dn_X/dt)(z=0)$, and the cosmological evolution of this rate,
determined by $p$.

The shapes of the UHE nucleon and $\gamma$-ray spectra predicted
within TD models are ``universal''
in the sense that they depend only on the
physics of X particle decay. This is because at UHEs 
nucleons and $\gamma$-rays
have attenuation lengths in the cosmic microwave background (CMB)
which are small compared to the Hubble scale. Cosmological evolutionary
effects which depend on the specific TD model are therefore negligible.
Since the resulting spectra tend to be harder than any other
components from acceleration sources, TD mechanisms could contribute
to the flux dominantly above $\simeq100\,$EeV, but negligibly
in the range $10^{14}\,{\rm eV}-10\,$EeV. 

In contrast to the universality of UHE spectral shapes, the
predicted  $\gamma$-ray flux below $\sim10^{14}\,$eV
(the threshold for pair production of photons on the CMB)
and the predicted neutrino flux 
depend on the total energy release integrated over
redshift and thus on the specific TD model. Compared to
acceleration scenarios, this energy release
can be substantial, especially at high redshifts
where conventional sources such as galaxies are not expected to
contribute. The production of UHE particles in TD scenarios is therefore
subject to a variety of constraints of mostly
cosmological nature. Electromagnetic (EM) energy
injected into the universe above the pair production threshold
on the CMB is recycled into a generic cascade spectrum below
this threshold on a time scale short compared to the Hubble
time. This can have several potentially observable
effects~\cite{SJSB} such
as modified light element abundances due to $^4$He
photodisintegration~\cite{Ellis} and distortions of the CMB.
Comparison with observational data
already rules out the class of TD scenarios corresponding to
$p=0$~\cite{SJSB} to which certain superconducting cosmic string models
belong~\cite{HSW}.

In addition, for a TD scenario to be viable,
its predicted spectrum must be consistent with all
flux measurements and limits available for
energies between $\simeq100\,$MeV and a few $100\,$EeV.
Observational data on the universal $\gamma$-ray background in
the $1-10\,$GeV region~\cite{CDKF}, to which the
generic cascade spectrum would contribute directly, turn out to
provide an important constraint. Since the UHE
$\gamma$-ray flux is especially sensitive to certain astrophysical
parameters such as the extragalactic magnetic field (EGMF), a reliable
calculation of the predicted spectral shapes requires numerical
methods. To this end we recently performed extensive numerical
simulations for the propagation of extragalactic nucleons,
$\gamma$-rays and electrons (positrons) in the energy range
$10^{8}\,{\rm eV}<E<10^{25}\,$eV, details of which
are given in Refs.~\cite{SLC,Lee}. The calculations take into
account all the 
relevant interactions with the (redshift dependent) universal
low energy photon background in the radio, microwave and
optical/infrared regime. For nucleons this includes
production of $e^{+-}$ pairs (for protons) and 
multiple pions. The $\gamma$-rays interact via production of single
and double $e^{+-}$ pairs, and electrons via inverse Compton scattering
and triplet pair production. In addition, electrons can suffer
significant synchrotron losses in 
EGMFs. We assume a flat universe, a Hubble constant of
$H_0=75\,{\rm km}\;{\rm sec}^{-1}{\rm Mpc}^{-1}$, and zero
cosmological constant.

Neutrino
fluxes within TD scenarios have been calculated before in the
literature: Ref.~\cite{Yoshida} contains a discussion of the
(unnormalized) predicted spectral shape and
Ref.~\cite{Brandenberger} computes the absolute flux predicted
by specific processes such as cusp evaporation on ordinary
cosmic strings. Ref.~\cite{BHS} and Ref.~\cite{YDJS} also calculate
absolute fluxes
within scenarios such as the one discussed above, with Ref.~\cite{YDJS}
discussing in detail neutrino propagation and experimental
issues related to the HiRes and the Telescope Array experiments.
However, none of these discussions take into account the cosmological 
constraints on TD models, in particular from the low energy $\gamma$-ray 
spectrum.

\begin{figure}[ht]
\centerline{\psfig{file=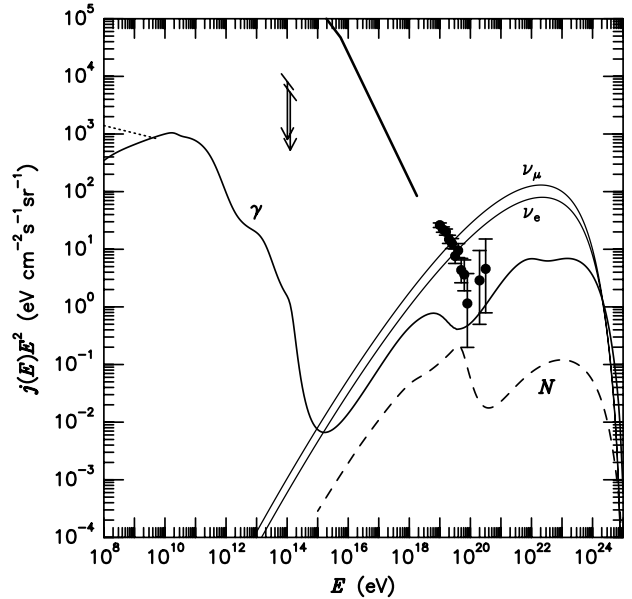,width=3.2in}}
\medskip
\caption[...]{Predictions for the differential fluxes of $\gamma$-rays (solid
line), nucleons (dashed line), and $\nu_\mu$
({\it or}\ $\bar\nu_\mu$), $\nu_e$ ({\it or} $\bar\nu_e$) 
(thin solid lines in decreasing order) by the TD model
characterized by $p=1$, $m_X = 2\times10^{25}\,$eV, assuming an
EGMF of $10^{-12}\,$G.
Also shown are the combined data from the Fly's
Eye~\cite{fe1} and the AGASA~\cite{ag1} experiments above $10\,$EeV
(dots with error bars), piecewise power law fits to the observed
charged CR flux (thick
solid line) and experimental upper limits on the $\gamma$-ray
flux at $1-10\,$GeV from EGRET data~\cite{CDKF} (dotted
line on left margin). The arrows indicate limits on the
$\gamma$-ray flux from Ref.~\cite{hegra}.
\label{F1}}
\end{figure}

Fig.~\ref{F1} shows flux predictions from our code for a typical
TD scenario for an EGMF of $10^{-12}\,$G. In this scenario, the
UHE CR events are assumed to be cascade
$\gamma$-rays produced by the decay of X particles whose
decay rate has been normalized accordingly.
Although $\gamma$-ray primaries might be somewhat disfavored~\cite{HVSV},
this is consistent with the currently unknown UHE CR
composition. The $\gamma$-ray and nucleon fluxes shown in
Fig.~\ref{F1} are consistent with observational estimates of the
integral flux above $300\,$EeV~\cite{fe1,ag1}, with a likelihood
significance above 50\% for $E\gtrsim100\,$EeV, and
with all data at lower energies.
Interestingly, the constraints arising from the diffuse
$\gamma$-ray background observed at $1-10\,$GeV~\cite{CDKF} are
somewhat less stringent than earlier analytical
estimates~\cite{Chi,SJSB}. As a consequence, for EGMFs
$\lesssim10^{-9}\,$G TD scenarios with X
particle masses as high as $10^{16}\,$GeV can still be viable
models of UHE CRs~\cite{SLC}. As in
Ref.~\cite{BHS}, the accompanying neutrino fluxes which are also
shown in Fig.~\ref{F1} were calculated using the absorption cross
section of UHE neutrinos in the thermal neutrino background. The
resulting fluxes
satisfy present bounds from the Fr\'{e}jus detector~\cite{Rhode}
within about five orders of magnitude. With fluxes of this
level, neutrinos are unlikely UHE CR candidates because of
their low interaction probability in the
atmosphere~\cite{SL}.

UHE neutrinos can produce muons in ordinary matter via charged
current reactions with nucleons (Ns)~\cite{FMR,GQRS}. The most recent
calculation of the corresponding cross section can roughly be
parametrized by $\sigma_{\nu N}(E)\simeq2.82\times10^{-32}\,{\rm
cm}^2\left(E/10\,{\rm EeV}\right)^{0.402}$ for $E\gtrsim1\,$PeV
where $E$ is the neutrino energy~\cite{GQRS}. We note that
uncertainties of the $\nu N$ cross sections above $10\,$EeV from
extrapolation of QCD evolution of up to a factor of 10 translate
into corresponding uncertainties in the predicted rates. For an (energy
dependent) ice or water equivalent acceptance $A(E)$ (in units
of volume times solid angle), one can obtain an approximate
expected rate of UHE muons produced by neutrinos with energy $>E$, $R(E)$, by
multiplying $A(E)\sigma_{\nu N}(E)n_{\rm H_2O}$ with the
integral muon neutrino flux $\simeq Ej_{\nu_\mu}$. Here,
$n_{\rm H_2O}$ is the nucleon density in water.
The neutrino energy and arrival direction can be reconstructed from
the observed muon bremsstrahlung and the track
geometry. Alternatively, one could use acoustic detection
methods. The backgrounds are in
general expected to be small~\cite{detector,Price}. The rate
prediction from the model shown in Fig.~\ref{F1} can be written as
\begin{equation}
  R(E)\simeq6\times10^{-3}\left[{A(E)\over1\,{\rm
  km}^3\times2\pi\,{\rm sr}}\right]\left({E\over10\,{\rm
  EeV}}\right)^{-0.1}\,{\rm yr}^{-1}\label{r1}
\end{equation}
around $E\sim10\,$EeV. Above $\simeq100\,$EeV the corresponding
fluxes would dominate all present model predictions for AGN
neutrino fluxes~\cite{AGN} as well as the flux of
``cosmogenic'' neutrinos produced by interactions of UHE CRs
with the universal photon background~\cite{HS,YT,YDJS}. Note that the $p=0$
model in Ref.~\cite{BHS} which implies neutrino event rates
about three orders of magnitude higher than Eq.~(\ref{r1}) is
ruled out because of overproduction of low energy
$\gamma$-rays~\cite{SLC}. We also note that
UHE neutrinos can produce lepton pairs on the thermal neutrino
background via the Glashow resonance~\cite{Weiler} whose decay
products in turn contain secondary neutrinos. As was shown in
Ref.~\cite{Yoshida}, for $m_X\gtrsim10^{24}\,$eV this effect,
which was not taken into account in our simulation, can increase
neutrino fluxes around $100\,$EeV by factors of a few which makes our
estimate Eq.~(\ref{r1}) conservatively low. Our flux estimates
are further reduced compared to Ref.~\cite{YDJS} for similar
scenarios by our normalization procedure which assures
consistency of predicted $\gamma$-ray and nucleon fluxes with
observational data at all energies.

For detectors based on the fluorescence technique such as the
HiRes~\cite{hires} and the Telescope
Array~\cite{tel}, the sensitivity to UHE neutrinos is often
expressed in terms of an effective aperture $a(E)$ which is
related to $A(E)$ by $a(E)=A(E)\sigma_{\nu N}(E)n_{\rm
H_2O}$. For the cross sections from Ref.~\cite{GQRS}, the
apertures given in Ref.~\cite{hires} for the HiRes correspond to
$A(E)\simeq3\,{\rm km}^3\times2\pi\,{\rm sr}$ for
$E\gtrsim10\,$EeV for muon neutrinos. The expected acceptance
of the proposed Pierre Auger project for horizontal
UHE neutrino induced events is $\simeq20\,{\rm km}^3\,{\rm sr}$
at $10\,$EeV and $\simeq200\,{\rm km}^3\,{\rm sr}$ at
$10^{23}\,$eV~\cite{Zas}. We conclude that detection of neutrino
fluxes predicted by scenarios such as the one shown
in Fig.~\ref{F1} requires running a detector of acceptance
$\gtrsim100\,{\rm km}^3\times2\pi\,{\rm sr}$ over a period of a few
years. Again, the backgrounds seem to be negligible~\cite{YDJS}.

A more model independent estimate for the average event rate
$R(E)$ can be made if the underlying
scenario is consistent with observational nucleon and
$\gamma$-ray fluxes and the bulk of the energy is released above
the pair production threshold on the CMB. Let us assume that the
ratio of energy injected into the neutrino versus EM channel is a
constant $r$. As in Fig.~\ref{F1}, cascading effectively reprocesses
most of the injected EM energy into low
energy photons whose spectrum peaks at $\simeq10\,$GeV~\cite{CA}.
Since the ratio $r$ remains roughly unchanged during
propagation, the height of the
corresponding peak in the neutrino spectrum should
roughly be $r$ times the height of the low-energy
$\gamma$-ray peak, i.e., we have the condition
$\max_E\left[E^2j_{\nu_\mu}(E)\right]\simeq
r\max_E\left[E^2j_\gamma(E)\right].$ Imposing the observational
upper limit on the diffuse $\gamma$-ray flux around $10\,$GeV
shown in Fig.~\ref{F1},
$\max_E\left[E^2j_{\nu_\mu}(E)\right] \lesssim
2\times10^3 r \,{\rm eV}{\rm cm}^{-2}{\rm sec}^{-1}{\rm
sr}^{-1}$, then bounds the average neutrino rate at all
energies $E \gtrsim1\,$PeV to be 
\begin{equation}
  R(E)\lesssim0.34\,r\left[{A(E)\over1\,{\rm
  km}^3\times2\pi\,{\rm sr}}\right]
  \,\left({E\over10\,{\rm EeV}}\right)^{-0.6}\,{\rm
  yr}^{-1}\,.\label{r2}
\end{equation}
For $r\lesssim10^5(10\,{\rm EeV}/E)$ this bound is consistent
with the constraint from the Fr\'{e}jus experiment~\cite{Rhode}.
In typical TD models such as the one discussed above,
$r\simeq0.3$. However, mechanisms with $r\gg1$ could induce
appreciable event rates above $\sim1\,$EeV in a km$^3$
scale detector. A detection would thus open the exciting
possibility to establish an experimental lower limit on $r$.
We stress that
Eq.~(\ref{r2}) holds regardless of whether or not the underlying
TD mechanism explains the observed UHE CR events.

The transient event rate could be much higher than Eq.~(\ref{r2}) in
the direction to discrete sources which emit particles in
bursts. Corresponding pulses in the UHE
nucleon and $\gamma$-ray fluxes would only occur for sources
nearer than $\simeq100\,$Mpc and, in case of protons, would be
delayed and dispersed by deflection in galactic and
extragalactic magnetic fields~\cite{MEW}. The recent observation
of a possible correlation of CRs above $\simeq40\,$EeV by the
AGASA experiment~\cite{ag2} might suggest
sources which burst on a time scale $t_b\ll1\,$yr.
A burst fluence of $\simeq r\left[A(E)/1\,{\rm km}^3\times2\pi\,{\rm
sr}\right](E/10\,{\rm EeV})^{-0.6}$ neutrinos within a time
$t_b$ could then be expected. Associated pulses
could also be observable in the ${\rm GeV}-{\rm TeV}$
$\gamma$-ray flux if the EGMF is smaller than
$\simeq10^{-15}\,$G in a significant fraction of extragalactic
space~\cite{WC}.

In conclusion, by using a new particle propagation code we have
given conservative estimates of the UHE
neutrino flux above $1\,$EeV which is predicted by a typical TD
type scenario of UHE CR origin.
We demonstrated that the constraint imposed by requiring that TD
scenarios do not overproduce the measured universal $\gamma$-ray
background at $1-10\,$GeV implies an upper
limit on these neutrino fluxes which only depends on the ratio $r$
of energy injected into the neutrino versus EM
channel, and not on any specific TD scenario or even a possible
connection to UHE CRs. For $r\gtrsim1,$ neutrino fluxes
near this upper limit are potentially
detectable by a km$^3$ scale neutrino observatory. 
A detection of the UHE neutrino flux might establish an
experimental lower limit on $r$ and thus
allow important insights into new fundamental physics near the
GUT scale. Due to the increase of the $\nu N$ cross section with
energy, neutrino event rates tend to decrease less strongly with
energy than UHE CR event rates and the spectral shape and cutoff of the
neutrino flux (and thus $m_X$) might be more easily accessible.
A non-detection with more stringent upper limits
would also be useful since it
could eliminate large classes of TD models of UHE CR origin. For
example, failing to detect neutrinos above $\simeq10\,$EeV with
an exposure $A\cdot t$ would rule out scenarios of the type shown in
Fig.~\ref{F1} for $r\gtrsim(100\,{\rm km}^3\,2\pi\,{\rm
sr}\,{\rm yr}/A\cdot t)$. Neutrino astronomy might
thus be connected to new fundamental physics.

We are grateful to Shigeru Yoshida for comments and
acknowledge useful discussions with Jim Cronin, Al Mann,
Chris Hill, Wolfgang Ochs, Paul Sommers, and Pijush Bhattacharjee.
This work was supported by the DoE, NSF and NASA at the University of Chicago,
by the DoE and by NASA through grant NAG5-2788 at Fermilab, and
by the Alexander-von-Humboldt Foundation. S.L. acknowledges the
support of the POSCO Scholarship Foundation in Korea. G.S.
thanks the Aspen Center for Physics for hospitality and
financial support.

\end{document}